\documentclass[conference]{IEEEtran}
\IEEEoverridecommandlockouts
\usepackage{cite}
\usepackage{amsmath,amssymb,amsfonts}
\usepackage{algorithmic}
\usepackage{graphicx}
\usepackage{textcomp}
\usepackage[table,x11names]{xcolor}
\usepackage{url}
\def\BibTeX{{\rm B\kern-.05em{\sc i\kern-.025em b}\kern-.08em
		T\kern-.1667em\lower.7ex\hbox{E}\kern-.125emX}}
\definecolor{upmaroon}{rgb}{0.5, 0.00, 0.00}

\usepackage{listings}
\usepackage{color}
\usepackage{tcolorbox}
\usepackage{tikz}
\usepackage{booktabs}
\usepackage{multirow}
\usepackage{verbatimbox}
\newcommand\blfootnote[1]{%
  \begingroup
  \renewcommand\thefootnote{}\footnote{#1}%
  \addtocounter{footnote}{-1}%
  \endgroup
}
\usepackage[colorlinks=true, urlcolor  = upmaroon, linkcolor=upmaroon, citecolor=blue]{hyperref}%

\definecolor{codegreen}{rgb}{0,0.6,0}
\definecolor{codegray}{rgb}{0.5,0.5,0.5}
\definecolor{codepurple}{rgb}{0.58,0,0.82}
\definecolor{backcolour}{rgb}{0.95,0.95,0.92}

\begin{document}
	\title{Coding Practices and Recommendations of Spring Security for Enterprise Applications} 
	
	\author{\IEEEauthorblockN{Mazharul Islam\IEEEauthorrefmark{1}, Sazzadur Rahaman\IEEEauthorrefmark{1}, Na Meng\IEEEauthorrefmark{1}, Behnaz Hassanshahi\IEEEauthorrefmark{2}, Padmanabhan Krishnan\IEEEauthorrefmark{2},\\ 
	Danfeng (Daphne) Yao\IEEEauthorrefmark{1}}
		\IEEEauthorblockN{Virginia Tech, Blacksburg, VA\IEEEauthorrefmark{1}, Oracle Labs, Australia\IEEEauthorrefmark{2}}
		
		\IEEEauthorblockN{\{mazharul, sazzad14, nm8247, danfeng\}@vt.edu, \{behnaz.hassanshahi, paddy.krishnan\}@oracle.com}
	}
	\maketitle

	\begin{abstract}
	  Spring security is tremendously popular among practitioners for its ease of use to secure enterprise applications.
	  In this paper, we study the application framework misconfiguration vulnerabilities in the light of Spring security, which is relatively understudied in the existing literature. 
		Towards that goal, we identify 6 types of security anti-patterns and 4 insecure vulnerable defaults by conducting a measurement-based approach on 28
		Spring applications. Our analysis shows that security risks associated with the identified security anti-patterns and insecure defaults can leave the enterprise application vulnerable to a wide range of high-risk attacks. 
		To prevent these high-risk attacks, we also provide recommendations for practitioners. 
		Consequently, our study has contributed one update to the official Spring security documentation while other security issues identified in this study are being considered for future major releases by Spring security community.

	\end{abstract}
	

	
	\section{Introduction}
	
	\blfootnote{This work has been supported by the National Science Foundation under Grant No. CNS-1929701.}
	
	Application frameworks enable reusing software designs at the architecture level by capturing the common abstractions of an application domain~\cite{DBLP:conf/icse/SchmidtB03}. Spring is the most popular application framework for enterprise Java applications~\cite{jrebelstat, stackoverflowstat}. Spring security~\cite{springsecurity} offers reusable authentication and authorization modules for enterprise applications that are written in Spring. It also provides default protection against common web-application related vulnerabilities, e.g., CSRF protection, including common security response headers. Spring security is highly customizable by design to enable seamless integration with various use-cases. Unfortunately, the abuse of such customization capabilities can be a great source of application insecurity. For example, in~\cite{Meng-ICSE-2018}, authors observed a common trend of disabling CSRF protection for convenience in StackOverflow posts. Without careful consideration, such customization can render a web application vulnerable to classic CSRF attacks.
	
	Prior research on security issues arising from reusable software components mostly focuses on library API misuse~\cite{DBLP:conf/ccs/GeorgievIJABS12,  DBLP:conf/ccs/FahlHMSBF12, egele2013empirical, DBLP:conf/sp/FischerBXSA0F17, DBLP:conf/ecoop/KrugerS0BM18, DBLP:conf/pldi/PaletovTRV18, rahaman2019cryptoguard}, while framework misconfigurations are largely unexplored. For example,~\cite{DBLP:conf/ccs/GeorgievIJABS12, DBLP:conf/sp/FischerBXSA0F17,Meng-ICSE-2018,DBLP:conf/pldi/PaletovTRV18} focus on understanding the nature of security API misuses.~\cite{DBLP:conf/ccs/GeorgievIJABS12, egele2013empirical} showed the dangers of misusing application-level SSL/TLS and cryptographic APIs. Researchers extensively studied the role of StackOverflow's misleading advices~\cite{DBLP:conf/sp/FischerBXSA0F17,Meng-ICSE-2018,DBLP:conf/icse/ChenFMWG19}, poor API designs~\cite{DBLP:conf/sp/Acar0FGKMS17}, lack of proper guidelines~\cite{DBLP:conf/secdev/AcarSWWMF17}, etc., behind this insecurity. Most of the existing methods to detect security API misuses rely on static code analysis~\cite{DBLP:conf/ccs/FahlHMSBF12, egele2013empirical, DBLP:conf/ecoop/KrugerS0BM18,DBLP:conf/ndss/BianchiFMKVCL18, rahaman2019cryptoguard}. Although, misconfiguring of security modules in application frameworks has great potential to cause insecurity, their nature, severity, prevalence, and detection feasibility are still mostly unknown. 
	
	In this paper, we present a thorough study of framework misconfiguration vulnerabilities in Spring Security. Our goal is to identify various classes of these vulnerabilities (referred to as \textit{security anti-patterns}), their nature, severity, and prevalence. Specifically, we pose the following research questions. \textit{What are the common security anti-patterns in enterprise Spring security applications? How severe are they? Most importantly, how prevalent are they in real-world enterprise software?}
	
	To find answers, we took a measurement-based approach. We manually analyzed 28 Spring-based applications hosted on GitHub to observe any insecure customization (i.e., security anti-patterns) of Spring security's i) \textit{authentication} and ii) \textit{protection against exploits} features. We also studied the security of the default configurations of these features. Our analysis discovered 6 types of security anti-patterns. We observed that programmers tend to intentionally disable CSRF protections, store secrets insecurely, use lifelong expiration of access tokens, etc. Our analysis of Spring security's default configuration revealed 4 major vulnerabilities. Our analysis found that Spring security uses 10 as the default strength ($2^{10}$ number of rounds) in Bcrypt during password encoding, while it is recommended to use at least $16$~\cite{rounds-bcrypt} to be secured. 
	We also found that Spring security uses insecure MD5 hashing to generate the ``remember me'' cookie. Most importantly, we identified that Spring security does not offer any throttling policy to limit the number of requests by users during API invocation. This insufficiency might lead to denial of service (DoS) attacks to applications using Spring security's OAuth functionality. 
	Our findings on 4 major vulnerabilities of Spring security's default configuration resulted in one update to Spring security's official documentation, while other issues are being considered for future major releases after we disclosed them to the Spring security community.
	
	In summary our contributions are as follows.
	
	\begin{itemize}
		\item Our analysis of 28 applications identified 6 common Spring security anti-patterns that undermine its security guarantees. During our analysis, we discovered 17 instances of disabling CSRF tokens, 14 instances of hard coded JWT signing key, 17 instances of storing secrets insecurely. 
		We also analyze the security risk associated with them and highlight recommendations for practitioners on how to avoid the anti-patterns and thus improve security.
		
		\item Our analysis of Spring security's default configuration revealed 4 major vulnerabilities, including the insecure use of Bcrypt for password encoding, the use of MD5 hash to generate ``remember-me'' cookie. We also identified that the lack of throttling policy per API key is susceptible to denial of service attacks.
		
		\item We divided our dataset into two groups, i.e., real-world applications (8) and demo (20) projects and cross-analyzed 6 security anti-patterns across the groups. Our analysis revealed that the anti-pattern's count ratio is higher in demo projects than the real-world applications. However, the nature of the anti-patterns is similar across the two group. 
		
	\end{itemize}



	\section{Threat Model and Methodology}
	\label{sec:threat-model-methodology}
	In this section, first, we present the threat model and then, discuss the methodology of our study.
	
	\subsection{Threat model}
	
	Existing cryptographic API misuse studies (e.g.,~\cite{DBLP:conf/ccs/GeorgievIJABS12,  DBLP:conf/ccs/FahlHMSBF12, egele2013empirical, DBLP:conf/sp/FischerBXSA0F17, DBLP:conf/ecoop/KrugerS0BM18, DBLP:conf/pldi/PaletovTRV18, rahaman2019cryptoguard}) are not specific for Spring security framework. Enterprise security issues in Spring security, such as the abuse of customization of reusable components or improper security management policies, are not well examined, with a few exceptions. For example, studies in~\cite{dikanski2012identification, Attribute-Based-Access-Control-for-APIs-in-Spring-Security} designed authentication and authorization patterns and access control policies of Spring security. In comparison, our paper aims to report security anti-patterns in Spring projects and their security threats. Specifically, we focus on the misconfigurations of two Spring security features, i.e., i) \textit{authentication} and ii) \textit{protection against exploits}. We also study the security status of default configurations of these features.
	
	\smallskip
	\noindent
	{\em \bf {Authentication.}} Spring security offers 9 types of authentications~\cite{authentication-mechanisms}. We analyze the use of 4 of these authentication mechanisms as follows, i) username-password, ii) ``remember me'' cookie, iii) OAuth 2.0 and iv) Java authentication and authorization service (JAAS)-based authentication. 
	Username-password based authentication is the most common way to authenticate users while ``remember me'' cookie facilitates remembering users between sessions. JASS and OAuth2.0 based authentication are a bit different since they delegate the authentication requests to their corresponding JASS server and OAuth 2.0 provider respectively. 
	
	Misconfiguring them can lead to a wide range of problems i.e., leaking application secrets (e.g., access tokens, passwords, etc), enabling man-in-the-middle (MitM), denial of service (Dos) attacks, etc.
	
	\smallskip
	\noindent
	{\em {\bf Protection against exploits.}} Spring security also provides protections against common exploits. In most of the cases, these protections are enabled by default. To protect against CSRF attacks, Spring security offers the following protections; i) CSRF token and ii) ``SameSite Attribute''-based protection. CSRF token-based protection ensures the presence of CSRF token in the HTTP request header to indicate its legitimacy. In ``SameSite Attribute''-based protection, the browser only sends the ``SameSite'' session cookie if and only if both the requested resource and the resource in the top-level browsing context match the cookie. Spring security also enables sending common HTTP security headers by default including, HTTP Strict Transport Security, X-Frame-Options, X-XSS-Protection, etc. Spring security also enables \textit{Strict Transport Security} by default to redirect HTTP traffic to HTTPS.

	\subsection{Methodology of the study}
	
	To systematically discover security anti-patterns, we develop the following methodology. First, we collect a dataset of real-world enterprise application source codes that uses Spring security. Then, we carefully analyze and collect their security configurations by using a descriptive coding technique~\cite{saldana2015coding}. After finding the use of a security feature, we extensively match its configuration with the following seven knowledge-base of common security issues; i) Common Weakness Enumeration (CWE)~\cite{mitre}, ii) Openstack security anti-patterns alert list~\cite{openstack}, iii) Spring security official reference guide~\cite{spring-security-reference}, iv)  Apigee Edge anti-patterns~\cite{apigee}, v) Snyk vulnerability Database~\cite{snyk-vuln-DB}, vi) previous research work on security anti-patterns~\cite{antiPattern-Nafess, rahman2019anti, siriwardena2014advanced, do-developers-update-their-libs, del2015analyzing}, vii) RFC documents~\cite{rfc6749-oauth2, rfc6819completeoauth, rfc7515, rfc8725}. If any of these knowledge-base indicates an insecure configuration, we analyze severity in the context of their usage. \textit{If an insecure configuration is the result of a customization, we mark it as a security anti-pattern. Otherwise, we mark it as an insecure default configuration.}


 	
 	\section{Security Anti-patterns in Spring Security}
 	\label{section-3}
 	
 	\begin{table*}[h]
 		\caption{Identified security misuses are presented with their corresponding knowledge-base references, affecting features, threats, severity and counts in 28 GitHub projects. High, medium, and low severity levels are denoted by H/M/L respectively.}\label{tab:Table1}
 		\vspace{-5pt}
 		\begin{center}
 			\begin{tabular}{@{}lllllll@{}}
 				\toprule
 				Type                               & Rule                                           & Reference & Feature            & Threat            & Severity & Count (28)\\ \midrule
 				\multirow{6}{*}{Anti-patterns}     & (1) Using lifelong valid access tokens         &    ~\cite{apigee, mitre, siriwardena2014advanced}       & Authentication     & Secrets leaking   & M   & 7  \\
 				& (2) Absence of state param in redirect URL     &   ~\cite{siriwardena2014advanced}        & Authentication     & CSRF attacks      & H  & 11 \\
 				& (3) Using fixed secret to sign JWT tokens      & ~\cite{mitre, rahman2019anti, rfc7515}          & Authentication     & Brute-force        & M & 14 \\
 				& (4) Storing secrets in insecure places         &  ~\cite{snyk-vuln-DB}         & Authentication     & Secrets leaking   & H  & 17 \\
 				& (5) Disabling CSRF protection                  &   ~\cite{snyk-vuln-DB, antiPattern-Nafess}        & Exploit protection & CSRF attacks      & H  & 17 \\
 				& (6) Not using TLS for HTTP communication       &   ~\cite{mitre, snyk-vuln-DB, rahman2019anti}        & Exploit protection & Man-in-the-middle & H & 15 \\
 				\cmidrule{1-7}
 				\multirow{4}{*}{Insecure defaults} & (7) Using Bcrypt with insecure params          & ~\cite{spring-security-reference}          & Authentication     & Brute-force        & H & 11 \\
 				& (8) Using MD5 in remember me cookie            &   ~\cite{openstack, rahman2019anti}        & Authentication     & Brute-force        & H & N/A\\
 				& (9) Lack of req. throttling policy per API key &    ~\cite{rfc6819completeoauth, mitre}       & Exploit protection & DoS attacks       & L & N/A\\
 				& (10) Absence of content security policy (CSP)  &  ~\cite{snyk-vuln-DB}         & Exploit protection & Code injection    & L  & N/A\\ \bottomrule 
 			\end{tabular}
 		\end{center}
 		\vspace{-15pt}
 	\end{table*}
 	
 	In this section, first, we will present our analysis result on the collected data and then briefly illustrate each common security misuse and their severity.
 	
 	\noindent
 	\textbf{Data collection.}  We collected the source code of 28 applications hosted on GitHub that uses Spring security. 8 of the selected projects are real-world enterprise applications and 20 of them are demo projects with example use of Spring security framework. We considered the following three criteria to filter them~\cite{munaiah2017curating}:
 	\begin{itemize}
 		\item \textbf{\#Forks.} The number of times the project has been forked. This gives an indication that these repositories have been adopted widely by other developers in their own code base \cite{jiang2017and}.  
 		\item \textbf{\#Stars.} The number of times the project has been starred by other developers which ensures that the curated repository is popular \cite{borges2016understanding} among other developers.
 		\item \textbf{Originality.} The project is not a clone or copy of another existing project.
 	\end{itemize}
 	
 	
 	\subsection{Analysis Result}
 	After analyzing the usage of Spring security framework of the selected projects, we identified 6 Spring security anti-patterns and 4 insecure default behaviors. Table~\ref{tab:Table1} presents these security misuses, with their reference knowledge-base, affecting features, threat, severity, and counts in 28 projects. After that, we divided the dataset into two groups i.e., i) 8 real-world applications and ii) 20 demo projects. Then we cross-checked the security misuse instances across them. Table~\ref{fig:frequency-analysis-demo-projects}, presents the results of our analysis. Although the misuse count ratio is higher in demo projects than the real-world projects, \textit{the nature of the misuse cases are vastly overlapped (Column 3 in Table~\ref{fig:frequency-analysis-demo-projects}).} 
 	It will be interesting to see whether (and how) developers are being influenced by these insecure demo codebases that can be directly copied.  
 	
 	Next, we describe each of them, their severity, and recommend suggestions on how developers can properly resolve them. After that, we present several interesting case studies.
 	
 	\begin{table}[h]
 		\caption{Security misuse counts for 8 real-world and 20 demo projects.} 
 		\label{fig:frequency-analysis-demo-projects}
 		\vspace{-10pt}
 		\begin{center}
 			\resizebox{\linewidth}{!}{
 				\begin{tabular}{@{}llll@{}}
 					\toprule
 					Anti-patterns                                           & Real-word   & Demo & Common\\
 					& projects (8)  & projects (20) & cases\\
 					\midrule
 					(1) lifelong valid access tokens         & 1 & 6  &  0   \\
 					(2) Absence of state param & 2 & 9 & 2\\ 
 					(3) Fixed secrets to sign JWT tokens & 6 & 8 & 4\\ 
 					(4) Storing secrets in insecure places & 5 & 12 & 5\\ 
 					(5) Disabling CSRF protection & 6 & 11 & 6 \\
 					(6) Not using TLS & 4 & 11 & 4\\
 					\bottomrule
 				\end{tabular}
 			}
 		\end{center}
 	\end{table}
 	
 	
 	
 	\subsection{Common Spring security anti-patterns}
 	\label{sec:security-antipatterns}
 	\subsubsection{Using lifelong valid access tokens}
 	Spring security allows the developers to specify an expiration time for each randomly generated access tokens. 
 	Developers want access token with a long lifetime as they are easier to manage. However, on the other hand, long lifetime increases the risk of reply attacks if any access token gets leaked.
 	The general advise from~\cite{apigee, openstack, rfc6749-oauth2} is to keep the life time just a bit longer than a normal user session time which can be generalized to a period of 15 minutes to 2 hours depending on different use cases. 
 	However, we have noticed a security anti-pattern among developers of setting lifetime of access token primarily arbitrary long in the range of 10-20 days as shown on listing~\ref{code:life-time}.
 	
 	\begin{lstlisting}[caption={Setting lifelong valid access token}, label={code:life-time}]
 	app:
 	auth:
 	|\includegraphics[scale=.015]{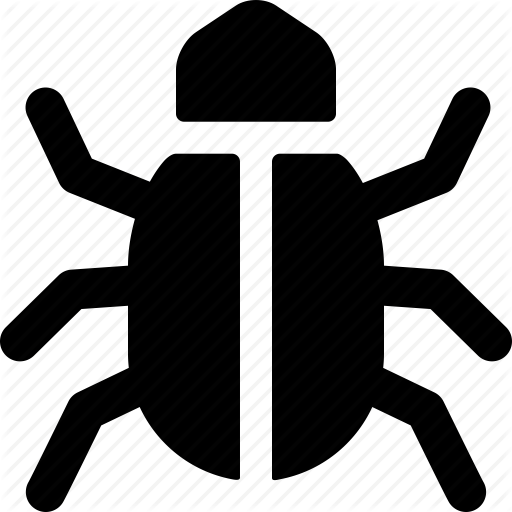}| tokenExpirationMsec: |\underline{864000000}|
 	// setting unnecessary long lifetime of 10 days
 	\end{lstlisting}
 	To avoid this security anti-pattern, we suggest the developers to minimize the lifetime of access token as much as possible so that whenever an attacker tries to reply the previously leaked secret access token it would already pass the expiration period. We also suggest to leverage refresh token to facilitate the user to provide a new access token effectively each time when the lifetime of an access token expires. 
 	
 	\subsubsection{Absence of \texttt{state} parameter in OAuth 2.0 Redirect URLs}
 	
\begin{figure}[h]
 		\centering
 		
 		\includegraphics[width=0.9\linewidth]{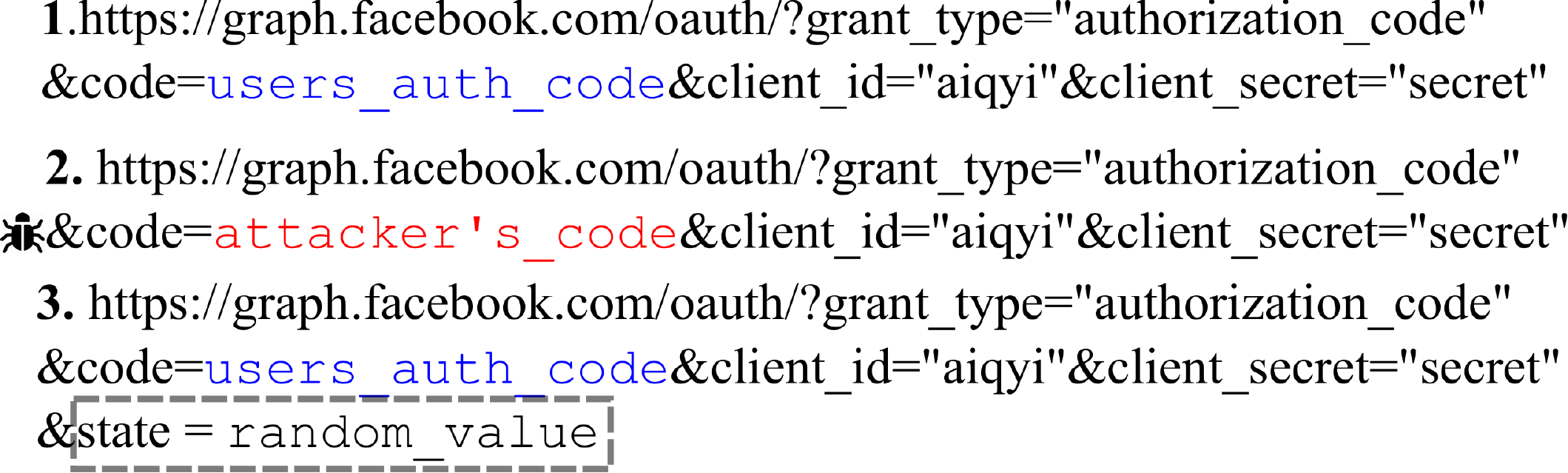}
 		\caption{The 2\textsuperscript{nd} redirect URL constructed from 1\textsuperscript{st} is vulnerable to forced login CSRF attack. The 3\textsuperscript{rd} redirect URL is not vulnerable due to the non guessable state parameter.}
 		\label{fig:csrf_redirectionURL}
 	\end{figure}

 	The continuous influx of increasing popularity of OAuth 2.0 \cite{oauth2.0} among developers motivated the active Spring security community to introduce the support for OAuth 2 in its lasted release~\cite{spring-security-5.X-new-features}. 
 	One of the most crucial parts among many of OAuth 2.0 authorization framework is sending the \texttt{auth\_code} generated by authorization server to client applications 
 	via redirection URLs~\cite{rfc6749-oauth2}. 
 	Interestingly redirect URL has a well define structure with guessable query parameters (1\textsuperscript{st} redirect URL on Fig.~\ref{fig:csrf_redirectionURL}). 
 	This can enable the attacker to construct a similar but malicious redirect URL by replacing user's auth code with their own auth code (2\textsuperscript{nd} redirect URL on Fig.~\ref{fig:csrf_redirectionURL}). By making the victim user clicking on this malicious redirect URL, attacker can perform a forced login CSRF attack~\cite{barth2008robust}. 
 	
 	To prevent this, RFC-6819~\cite{rfc6819completeoauth} has recommended a strict guideline to add an additional \texttt{state} parameter - the value of which is randomly generated (as shown on the 3\textsuperscript{rd} redirect URL in Fig.~\ref{fig:csrf_redirectionURL}). In this way the attacker won't be able to guess the value of state parameter and construct a malicious redirect URL. 
 	However, we have noticed in contrast to this strict recommendation, 
 	developers tend not to use the additional \texttt{state} parameter in redirect URL rendering 
 	the client application  vulnerable to previously mentioned forced login attack CSRF attack.  Hence we consider the missing of \texttt{state} parameter in redirect URL as a security anti-pattern. 
 	
 	The proper way to handle this security anti-pattern is to randomly generate a value and add this value to the state parameter in the redirect URL as shown on listing~\ref{code:state-missing}.
 	\begin{lstlisting}[caption={Adding \texttt{state} param in \texttt{redirect\_URL}}, label={code:state-missing}]
 	public String getToken(@RequestParam String code){
 	...
 	params.add("grant_type","authorization_code");
 	params.add("code",code);
 	params.add("client_id","aiqiyi");
 	params.add("client_secret","secret");
 	|\includegraphics[scale=.05]{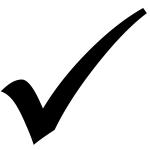}| |\hspace{0.3cm}|        params.add(|\underline{"state",127621437303857}|);
 	// Randomly generated value of state param
 	...
 	}
 	\end{lstlisting}
 	
 	
 	\subsubsection{Using fixed secrets to sign JWT tokens}
 	Spring security facilities the use of JSON Web Tokens (JWT)~\cite{rfc7519jwt} to authenticate users by 
 	adding \texttt{JwtTokenFilter} to the \texttt{DefaultSecurityFilterChain} with only a few lines of additional code. The way JWT works is that there a set of claims embedded inside JWT and the server signs these claims cryptographically using secret key(s). The user must present these cryptographically signed claims to the server and then the server verifies them to check  the authentication of these presented claims.
 	This design allows the server to be stateless and consequently scalable
 	which is one of the major reasons behind the emerging popularity of JWT.
 	
 	However, we noticed that developers tend to sign the claims of JWT cryptographically by predictable fixed secret key which makes it inherently vulnerable \cite{egele2013empirical} and relatively easy for attackers to crack the fixed secret key by brute force attacks.
 	\begin{lstlisting}
 	public class TokenProvider {
 	public String createToken(Authentication auth) {
 	Public String JWT_SIGN_KEY = "123456";
 	token = Jwts.builder()
 	...
 	|\includegraphics[scale=.015]{bug.png}| |\hspace{1 cm}|              .signWith(SignatureAlgorithm.HS512, |\underline{JWT\_SIGN\_KEY}|)
 	// signing with a hard coded fixed secret 
 	.compact();
 	...
 	return token; 
 	}
 	}
 	\end{lstlisting}
 	This insecure practice of using a fixed secret to sign all JWT can lead to a great inconvenience frequently when any one of the valid JWT gets leaked. Because then invalidating the leaked JWT would require the developer to change the fixed secret manually. This would automatically invalidate all other valid JWT as well which have not been leaked. Perhaps a quick way around instead of changing the fixed secret would be to keep a blacklist for leaked JWT in the database. However keeping, maintaining, and querying to the database if the JWT is blacklisted for each request can be expensive and most importantly introduces scalability problems, which is one of the very basic problems why JWT was introduced and is popular among developers.
 	
 	RFC-8725 \cite{rfc8725} strongly discourages this insecure coding practice and advises the developers to use random cryptographic keys with sufficient entropy. To solve this problem, we suggest to leverage the optionally available JWT key identifier parameter \texttt{kid} which can be leveraged for managing multiple randomly generated secret keys~\cite{rfc7515}. 
 	
 	\subsubsection{Storing secrets in insecure places}
 	To avoid application specific secrets e.g., password, DB-root password from being leaked developers need to avoid keeping them in unsafe places (e.g., plain-text, local storage). We have noticed a security anti-pattern among developers to write the application specific secrets to \texttt{application.yml} configuration file in plain text as shown in listing~\ref{code:application.yml}. 
 	
 	\begin{lstlisting}[caption={Stroing secrets insecurely in \texttt{application.yml}}, label={code:application.yml}]
 	spring:
 	datasource:
 	username: root
 	|\hspace{1 cm}| |\includegraphics[scale=.015]{bug.png}|         password: |\underline{uOtmALFgsfxgYzEg1uLXl3O}|
 	\end{lstlisting}
 	To circumvent this security anti-pattern and to handle application specific secrets securely, Spring vault facilitates a succinct solution. Spring vault which is an abstraction layer on top of HashiCorp vault~\cite{vaultproject} providing annotation based access for clients to store and retrieve secrets in a secure way as shown in listing~\ref{code:storing-secrets}. 
 	
 	\begin{lstlisting}[language=java, caption={Retrieving secrets from Spring vault}, label={code:storing-secrets}]
 	@Value("${clientPassword}")
 	String client_password;
 	\end{lstlisting}
 	
 	
 	\subsubsection{Disabling CSRF protection}
 	\label{SAP:1}
 	
 	Spring security by default secures 
 	the web applications by defining a method \texttt{csrf()} and  implicitly enabling this function invocation for each state-changing request (e.g., PATCH, POST, PUT, DELETE)~\cite{stackoverflow-csrf}. We have observed a recurring insecure practice among developers to manually disabling the default CSRF protection as shown on listing~\ref{csrf-disable}. 
 	\begin{lstlisting}[language=java, caption={Manually disabling default CSRF protection},label={csrf-disable}]
 	@Override
 	protected void configure(HttpSecurity hs) throws Exception {
 	|\includegraphics[scale=.015]{bug.png}| |\hspace{2cm}|	            hs.csrf.|\underline{disable()}| 
 	}
 	\end{lstlisting}
 	Spring security's inherently different CSRF protection mechanism is a plausible answer to the prominence of this insecure coding practice. Spring security makes the CSRF token inaccessible to the front-end of the application 
 	by setting \texttt{httpOnly} to true. 
 	Being unable to access the CSRF token, front-end JavaScript frameworks (e.g., Angular JS, Vue JS, Laravel etc.) will throw configuration errors. To avoid these errors developers primarily tend to turn off the default CSRF protection the developers seeking a short workaround way without understanding the insecurity associate with doing so. 
 	
 	The proper way to circumvent this error is to 
 	
 	set  \texttt{httpOnly} to false as shown on listing~\ref{code:csrf-correct}. That being said, we want to point out that even though setting the \texttt{httponly} to false can reveal the CSRF token to JavaScript loaded from the same domain, third party JavaScript loaded from different domain will not be able to access the access token because of same-origin policy; thereby defeating Cross Site Scripting attacks.
 	
 	\begin{lstlisting}[language=java, caption={Proper way to circumvent CSRF misconfiguration errors},label={code:csrf-correct}]
 	@Override
 	protected void configure(HttpSecurity hs) throws Exception {
 	hs.csrf(c -> c
 	.csrfTokenRepository(CookieCsrfTokenRepository
 	|\includegraphics[scale=.05]{correct.png}|  |\hspace{2cm}|	                .|\underline{withHttpOnlyFalse()}|) 
 	// Marking the CSRF Token's HttpOnly to False.
 	)
 	}
 	\end{lstlisting}
 	We further want to emphasize that disabling  CSRF protection itself is not \textit{always} a security problem when authentication is done via bearer tokens. However, it can be a severe security problem when Spring security will perform authentication based on the authentication cookie (i.e., JESSIONID) stored on the user's browser. 
 	
 	
 	
 	\subsubsection{Not using TLS for HTTP communications} 
 	The security of many components of authentication and authorization packages in Spring security  (e.g., oauth 2.0, SAML 2.0, CAS, OpenID connect) recommend, and in some cases mandate the use of TLS. 
 	However, our analysis found that developers tend to avoid the use of TLS in many places and show a similar trend highlighted in previous study~\cite{Meng-ICSE-2018, rahaman2019cryptoguard, rahman2019seven}. 
 	\begin{lstlisting}
 	eureka:
 	client:
 	serviceUrl:
 	|\includegraphics[scale=.015]{bug.png}| defaultZone: |\underline{http}|:root@paascloud-eureka:8761
 	// use of HTTP without TLS
 	\end{lstlisting}
 	Although, it is difficult to create, install, find, and validate TLS certificates in development environment, we suggest the developers to enable TLS in the production environment as suggested in~\cite{DBLP:conf/ccs/GeorgievIJABS12}.
 	
 	
 	
 	
 	
 	\subsection{Insecure default configuration}
 	\label{insecure-defaults}
 	\subsubsection{Using BCrypt with insecure params}
 	
 	Spring security supports multiple ways to implement a \texttt{PasswordEncoder}. A brief summary of them is presented in section~\ref{appendix:password-hashing}. 
 	Here, we focus on \texttt{BCryptPasswordEncoder}, which is popularly used to encode passwords. 
 	We found that the default configuration of \texttt{BCryptPasswordEncoder} is vulnerable to brute-force attacks.

 	\texttt{BCryptPasswordEncoder} is based a deliberately slow hashing function Bcrypt~\cite{bcrypt}. 
 	This slowness or number of rounds in Bcrypt is one of the key factors which 
 	makes it resistant to do feasible brute-force attack. 
 	The security strength (i.e., slowness or number of rounds) of \texttt{BCryptPasswordEncoder} can be specified by the developers 
 	as a parameter to the constructor. However we noticed that the default strength of \texttt{BCryptPasswordEncoder} is $10$ ($2^{10}$ number of rounds)~\cite{bcrypt-spring-documentation}.
 	However as mentioned previously~\cite{rounds-bcrypt} and confirmed by our own experiments presented in Appendix~\ref{bcrypt-password-encoder},
 	this default strength does not have enough slowness and essentially lack the security strength needed to prevent brute-force attack.
 	
 	As developers tend not to pass any security strength parameter to the constructor assuming the default strength is secure enough, this tendency (as shown on listing~\ref{code:bcrypt}) leaves an exploitable opportunity for the attackers.
 	
 	\begin{lstlisting}[caption={Default strength is vulnerable to brute-force attack},label={code:bcrypt}]
 	@Bean
 	public PasswordEncoder passwordEncoder() {
 	|\includegraphics[scale=.015]{bug.png}| return new |\underline{BCryptPasswordEncoder()}|;
 	/* using default strength 10 is vulnerable 
 	to feasible brute-force attacks */
 	}
 	\end{lstlisting}
 	
 	Hence, we consider using the default insecure strength for \texttt{BcryptPasswordEncoder} as an insecure default configuration and recommend the developers to override the default strength by adjusting it according to their own system.
 	
 	
 	\subsubsection{Using weak hash algorithm MD5 in \texttt{remember-me} cookie}
 	Spring security provides \texttt{remember-me} which  stored in the browser. 
 	allows the browser to remember the user for future sessions and causing automated login to take place. This \texttt{remember-me} cookie is constructed from the  MD5 hashing of the username, expiration time of the cookie, and password and secret key as shown in listing~\ref{code:remember-me-cookie}.
 	
 	\begin{lstlisting}[caption={Construction of \texttt{remember-me} cookie}, label={code:remember-me-cookie}]
 	base64(username + ":" + expirationTime + ":" + 
 	|\includegraphics[scale=.015]{bug.png}|     |\underline{md5Hex}|(username + ":" + expirationTime +
 	// Use of weak hashing algorithm MD5 
 	":" password + ":" + key)) 
 	
 	\end{lstlisting}
 	The problem with this approach is that MD5 is considered broken, which susceptible to collision attacks~\cite{den1993collisions} and modular differential attacks~\cite{wang2005break}. Hence, attackers can easily recover sensitive information or temper the \texttt{remember-me} cookie. Therefore, we suggest the Spring security maintainers to fix this issue by replacing MD5 with a secure hashing algorithm (e.g., SHA-256).
 	
 	\subsubsection{Lack of required throttling policy per API key}
 	One of the most important parts of resource management policy for web API is to set a proper throttling policy per user. 
 	This throttling policy places a limit on the number of requests a user can make with a secret API key. 
 	Otherwise, an attacker can use use a valid secret API key to generate a massive number of requests than the web service can handle. In this way, the attacker will be able to make a denial of service attack for other users. 
 	However, unlike other security framework (e.g., Django~\cite{django-throttling}), Spring security lacks this as a built-in feature given the prevalence of this DoS attack with custom made bots. An IP-based throttling policy can prevent DoS attack but then attackers can switch to DDoS attack to abuse valid API keys.
 	We suggest a manual solution where the developers build custom filter and register it in the Spring context.
 	
 	\begin{lstlisting}[caption={Example of adding content security policy headers},label={code:content-security-policy}]
 	@Override
 	protected void configure(HttpSecurity hs) {
 	hs
 	...
 	.headers(headers -> headers
 	.contentSecurityPolicy(csp -> csp       
 	.policyDirectives("script-src 'self'"))
 	...
 	} 
 	\end{lstlisting}
 	
 	\subsubsection{Absence of content security policy}
 	Unexpectedly, unlike other protection mechanisms, Spring security does not add content security policy (CSP) HTTP headers by default. CSP helps the developers to enforce a fine-grained security policy easily to prevent code injection attacks e.g., cross site scripting, clickjacking, and data injection, etc. 
 	For example, if a developer perceives JavaScript from all external sources as untrusted then the developer needs to set the CSP header to \texttt{self} to prevent the browser from loading unsafe JavaScript from any untrusted external sources as shown on listing~\ref{code:content-security-policy}.
 	
 	We have noticed a prominent tend among developers of not adding the CSP headers manually even though adding the headers CSP to prevent varieties of code injection attacks is important. Especially, because these code injection attacks are not trivial to prevent from security stand point. Developers might perceive that just like other protection mechanisms CSP headers are provided by default on in Spring security. Hence, we consider the case of not setting the CSP HTTP headers as an insecure default.
 	
 	\subsection{Severity of the misconfigurations}
 	\label{sec:security-severity}
 	Each security  misuses discussed in the previous section~\ref{sec:security-antipatterns}, ~\ref{insecure-defaults} has specific attack vectors presented in the literature. To prioritize them based on their security severity, we group them into  three categories high, medium, and low. In this regard we consider two criteria i) attack difficulty, and ii) attacker’s gain as inspired from the Common Vulnerability Scoring System (CVSS) calculator~\cite{cvss}.
 	The assignments of these severities to identified security misuses are highlighted in Table~\ref{tab:Table1} and described as follows.
 	\subsubsection{High} Anti-patterns causing CSRF and man-in-the-middle attacks are easy to construct and provide large gain for the attackers. Insecure default MD5 in remember me cookie and BCrypt with insecure param can be brute-force easily given the high number of available cracking tools. Attacks originating from exploiting secrets stored in insecure places are easy to construct as well.
 	\subsubsection{Medium} Attackers can try to do offline brute force attack to guess the fixed secret key used to sign the JWT token which can be time consuming depending on the entropy/randomness level and length of the fixed secret key. Moreover lifelong access token can be reused/replayed by attackers if and only if  it is leaked. 
 	\subsubsection{Low} We place the two insecure defaults which depend on the presence of another vulnerability for the attacker to take advantage. For example absence of content security policy can be leveraged if and only if there is Cross site vulnerability already present. In the same way lack of required throttling policy per API key can cause DoS/DDoS attack if attacker can bypass the network level protection to mitigate DoS/DDoS attack in the first place. 
 	
 	

  
  
  
  \subsection{Case Study}
  \label{sec:case-study}
  
  
  \noindent
  \textbf{Simplicity and performance over insecure practices.}  
  In one real-world application, we saw the following comments before one of the security anti-pattern.
  
  \textit{``THIS IS NOT A SECURE PRACTICE! For simplicity, we are storing a static key here.''}
  
  In another case, a developer responded to us for a potential CSRF attack due to the absence of \texttt{state} parameter in redirect URL as following,
  
  \textit{``Increasing the state parameter can effectively prevent CSRF attacks. But my demo is just a simple sso demonstration. The simplest way to demonstrate the entire sso interaction process does not need to consider CSRF attacks.''}
  
  In another project, we noticed that the developer intentionally downgraded the default strength $10$ ($2^{10}$ rounds) of \texttt{BCryptPasswordEncoder} to $8$ to increase the performance. This illustrates that developers more often prefer simplicity and performance over secure coding practices.

  \noindent 
  \textbf{Storing application secrets in config file.} We observed that in 
  3 applications and 
  8 demo applications, developers stored secrets in the \texttt{application.yml}. Especially, we observed  
  5 number of applications to store secrets to sign the JWT in \texttt{application.yml} files. It is recommended to store these secrets in the server's key stores~\cite{rahaman2019cryptoguard}.
  
  \noindent
  \textbf{Separating development and production environments.} Sometimes developers avoid configuring TLS in their development environment~\cite{rahman2019seven}. A similar configuration will cause a devastating effect on the production environment. Spring security enables separating two different environments by simply using two separate configuration files. 
  However, we observed that in 13 projects, developers used the same configuration of insecure TLS in both development and production.
  
  
  
  \section{Discussion}\label{discussion}
  
  \noindent 
  {\bf Disclosure of our findings.}  We made an effort to share our concerns about these 3 major vulnerabilities found with the Spring security community. We created two pull requests and one issue on the master branch of Spring security project about i) replacing weak MD5 with a strong hashing algorithm SHA-256 ii) adding proper guidelines in Spring security official documentation about setting a secure strength for \texttt{BCryptPasswordEncoder} iii) possibility of performing DoS attack exploiting lack of required throttling policy per API key. The second request was already accepted and the community agreed with us on others. However, they expressed to remain \emph{passive} for now but will consider bringing our suggested changes in the next Spring security major release. We also reported some of the severe cases to the application developers and in the process of disclosing others.

  \noindent 
  {\bf Limitations.}
  The derived security anti-patterns are mainly based on manual inspection and therefore is subjected to human bias. To address this, the first two authors of the paper carefully and independently apply analysis multiple times to  verify the security anti-patterns. 
  We also acknowledge that the data-set constructed by popularity and adaption measure is susceptible to subjectivity, as this filtering measure may incorrectly remove some relevant projects using spring security. 


\section{Conclusion}
\label{sec:conclusion}

Without careful considerations, customizing application frameworks can cause critical vulnerabilities in an enterprise application. In this paper, we studied the application framework misconfiguration vulnerabilities in the light of Spring security. First, by analyzing 28 Spring security applications, we identified 6 security anti-patterns and 4 insecure default behaviors representing possible insecure use-cases of Spring security. Our analysis showed that the security anti-patterns are prevalent and similar across the real-world and demo applications, hence, pose a realistic threat.
	
	
	\bibliographystyle{ieeetr}
	\bibliography{main}
	\appendices

	\section{Implementations of Password hashing}
	\label{appendix:password-hashing}
	Spring security provides an array of \texttt{PasswordEncoder} implementation for storing password. We provide here a summary of these implementations.
	However among them \texttt{Md4Password}, \texttt{MessageDigest}, \texttt{Standard}, and \texttt{LdapSha} password encoders use digest based password encoding which is not considered secure. As a result they are  deprecated to indicate that they are legacy implementation. \texttt{Argon2Password} and \texttt{SCryptPassword} password encoder uses Bouncy castle. One problem associated with this is that Bouncy castle does not exploit parallelism/optimizations that other password crackers will. Therefore there is an uneven asymmetry between attacker and defender. The most recommended way to implement the \texttt{PassworEncoder} interface is \texttt{BCryptPasswordEncoder}.
	Table~\ref{tab:password-hashing} summarizes the 9 option for password encoding offered by Spring security.
	\begin{table}[h]
		\renewcommand{\arraystretch}{1.3}
		\caption{Summary of Password encoding interface implementations}
		\label{tab:password-hashing}
		\resizebox{\linewidth}{!}{
			
			\begin{tabular}{|l||l|}
				\hline
				\rowcolor{lightgray} \bfseries Implementation Name & \bfseries Comment\\
				\hline\hline
				\texttt{BCryptPasswordEncoder} & preferred implementation \\ \hline 
				\texttt{NoOpPasswordEncoder} & stores password in plain-text \\
				\hline 
				\texttt{Md4PasswordEncoder} & digest based password encoding\\
				\texttt{MessageDigestPasswordEncoder} & \\ 
				\texttt{StandardPasswordEncoder} & \\
				\texttt{LdapShaPasswordEncoder} & \\ 
				\hline
				\texttt{Argon2PasswordEncoder} & uses Bouncy castle\\ 
				\texttt{SCryptPasswordEncoder} & \\  
				\hline 
				\texttt{Pbkdf2PasswordEncoder} &  uses PBKDF2 \\ 
				\hline 
			\end{tabular}
		}
	\end{table}
	
	\section{Insecure default strength for \texttt{BCryptPassword Encoder}}
	\label{bcrypt-password-encoder}
	The Spring security reference guide mentions that the  
	strength of \texttt{BCryptPasswordEncoder} should be tuned to take about 1 second to verify a password on the developer's own system.
	However, according to our experiments the default strength of \texttt{BCryptPasswordEncoder} which is 10 (number of rounds $2^{10}$) takes around $~0.1$ seconds to verify a password. This is way less than 1 second lower limit. Since developers tend to use the default strength without any consideration assuming that the default implementation should have enough strength we consider this an insecure default configuration.
	
\begin{figure}[h]
		\centering
		\includegraphics[width=0.9\linewidth]{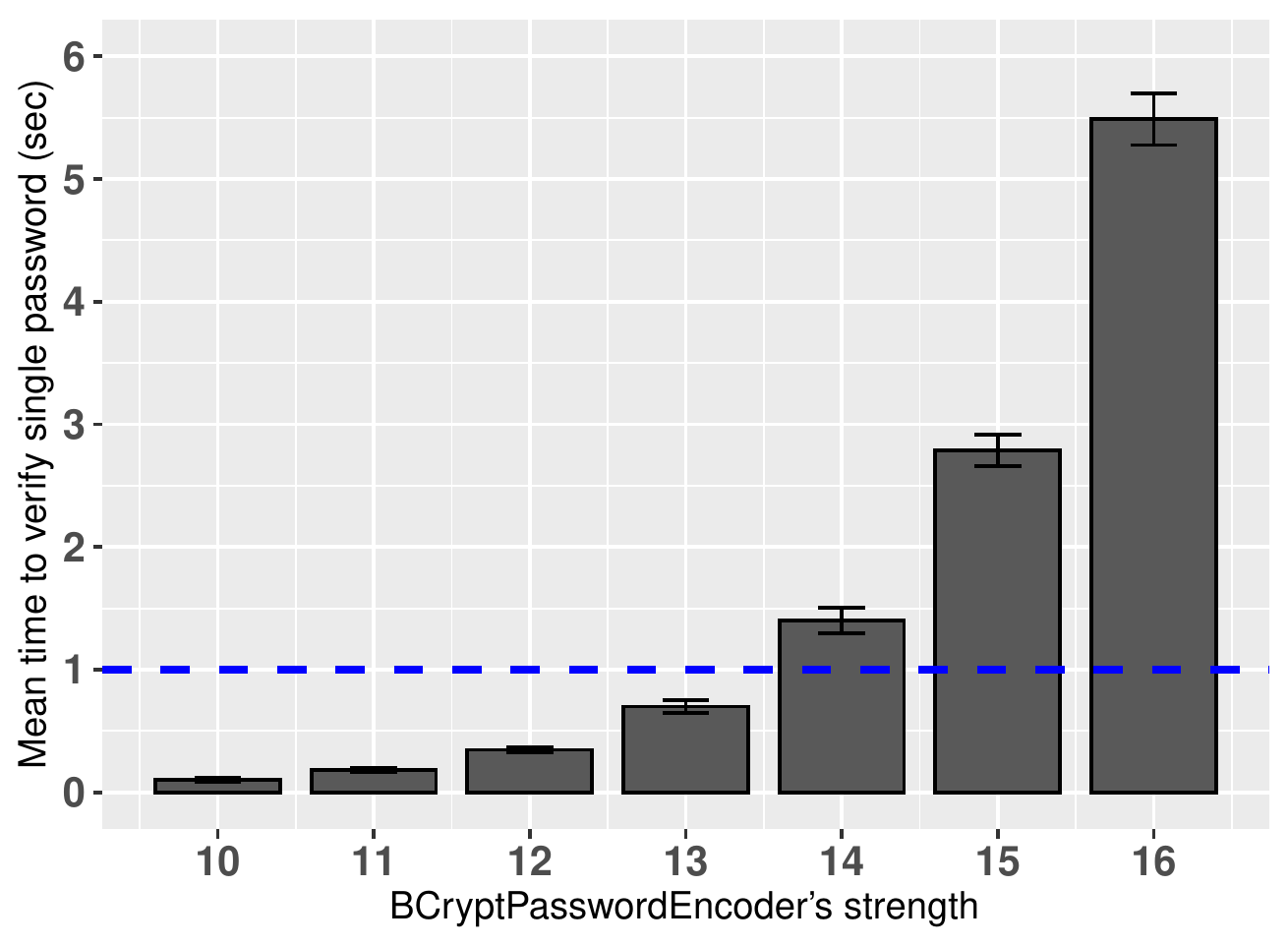}
		\caption{Mean time to verify a password of size 128 bits on our own system for different strength. Ideally the strength should add enough slowness so that it takes at least 1 sec to verify a password as marked by the horizontal dashed line.}
		\label{fig:encoding-time}
	\end{figure}

	\noindent
	On our system (LENOVO Ideapad- Intel(R) Core(TM) i5 CPU @ 1.60GHz - 7.4 GiB RAM - running Ubuntu 20.04 64 bit), we found out that appropriate secure strength should be 14 as shown on Figure~\ref{fig:encoding-time}. We emphasize that for GPU which can perform orders of magnitude faster than a typical CPU like the one we have used, the appropriate secure strength to prevent feasible brute-force attack should be higher than 14.

	\section{Finding vulnerabilities in 8 real world projects}
	\label{appendix:vulnerable-dependencies-test}
	We also test the 8 real-world projects using a know vulnerability scanning tool Snyk~\cite{snyk-vuln-DB} and then parse the scanning results. Snyk maintains an online database of known vulnerability and can automatically build the project to check against these known vulnerabilities. A summary of the results is shown in Table~\ref{tab:vul-dependencies}.
	\begin{table}[ht]
		\caption{\# of vulnerability found and their risk level in 8 real world applications using Snyk. Risk level High, medium, and low are denoted by H, M, L respectively.}
		\renewcommand{\arraystretch}{1.3}
		\resizebox{\linewidth}{!}{
			\begin{tabular}{|c|l|c|c|c|c|c|}
				\hline \hline
				\rowcolor{lightgray} \bfseries No & \bfseries Project & \bfseries \# forks & \bfseries \# stars & \bfseries LoC & \bfseries Risk  & \bfseries \# of \\ 
				\rowcolor{lightgray} & \bfseries name & & & & \bfseries level & \bfseries vulnerabilities\\
				\hline \hline 
				1 & paascloud-master  & 3.6k & 7.7k & 55.8k & H &  68  \\ \cline{6-7}
				& & & & &  M & 18 \\ \cline{6-7}
				& & & & & L & 7 \\ \hline 
				2 & xboot & 968 & 2.6k & 21.2k & H & 3 \\ \cline{6-7}
				& & & & & M & 3 \\ \cline{6-7}
				& & & & & L & 0 \\ \hline 
				3 & Spring-boot-cloud & 1.2k & 2k & 523 & H & 71 \\ \cline{6-7} 
				& & & & & M & 18 \\ \cline{6-7}
				& & & & & L & 7 \\ \hline 
				4 & sso & 323& 702 & 3.3k & H & 55\\ \cline{6-7}
				& & & & & M & 10 \\ \cline{6-7}
				& & & & & L & 5 \\ \hline 
				5 & FEBS-cloud & 326& 660 & 11.9k & H & 5\\ \cline{6-7}
				& & & & & M & 6 \\ \cline{6-7}
				& & & & & L & 1 \\ \hline 
				6 & fw-cloud-framework  & 325& 638 & 13.9k & H & 63 \\ \cline{6-7}
				& & & & & M & 6 \\ \cline{6-7}
				& & & & & L & 3 \\ \hline
				7 & cas &3.2k & 7.5k & 33.5k & \multicolumn{2}{l|}{No vulnerability found} \\ \hline
				8 & microservices-platform & 736& 1.6k & 25.4k &\multicolumn{2}{l|}{Can not run Snyk because of build errors} \\ 
				\hline 
			\end{tabular}
		}
		\label{tab:vul-dependencies}
	\end{table}

\end{document}